# Exciton Brightening in Monolayer Phosphorene via Dimensionality Modification

Renjing Xu,[1,†] Jiong Yang,[1,†] Ye Win Myint,[1] Jiajie Pei, [1,3] Han Yan,[1] Fan Wang,[2] and Yuerui Lu[1*]

[1]Research School of Engineering, College of Engineering and Computer Science, the Australian National University, Canberra, ACT, 2601, Australia

[2]ARC Centre for Nanoscale BioPhotonics (CNBP), Department of Physics and Astronomy Faculty of Science, Macquarie University, Sydney, NSW, 2109, Australia

[3]School of Mechanical Engineering, Beijing Institute of Technology, Beijing, 100081, China

[†] These authors contributed equally to this work

[*] To whom correspondence should be addressed: Yuerui Lu (yuerui.lu@anu.edu.au)

**Abstract**

**Two-dimensional (2D) monolayer phosphorene, a 2D system with quasi-one-dimensional (quasi-1D) excitons, provides a unique 2D platform for investigating the dynamics of excitons in reduced dimensions and fundamental many-body interactions. However, on the other hand, the quasi-1D excitonic nature can limit the luminescence quantum yield significantly. Here, we report exciton brightening in monolayer phosphorene achieved via the dimensionality modification of excitons from quasi-1D to zero-dimensional (0D), through the transference of monolayer phosphorene samples onto defect-rich oxide substrate deposited by plasma-enhanced chemical vapor deposition (PECVD). The resultant interfacial luminescent local states lead to exciton localization and trigger efficient photon emissions at a new wavelength of ~920 nm. The luminescence quantum yield of 0D-like localized excitons is measured to be at least 33.6 times larger than that of**

**intrinsic quasi-1D free excitons in monolayer phosphorene. This is primarily due to the reduction of non-radiative decay rate and the possibly enhanced radiative recombination probability. Owing to the large trapping energy, this new photon emission from the localized excitons in monolayer phosphorene can be observed at elevated temperature, which contrasts markedly with defect-induced photon emission from transition metal dichalcogenide (TMD) semiconductor monolayers that can only be observed at cryogenic temperatures. Our findings introduce new avenues for the development of novel photonic devices based on monolayer phosphorene, such as near-infrared lighting devices that are operable at elevated temperature. More importantly, 2D phosphorene with quasi-1D free excitons and 0D-like localized excitons provides a unique platform to investigate the fundamental phenomena in the ideal 2D-1D-0D hybrid system.**

## Introduction

Phosphorene, a newly discovered two-dimensional (2D) layered material, has aroused considerable interest because of its unique anisotropy[1-8], layer-dependent photoluminescence[9, 10], and quasi-1D excitons[11, 12]. All these properties are in stark contrast with the characteristics of graphene[13] and transition metal dichalcogenide (TMD) semiconductors[14-17]. In particular, 2D phosphorene with a quasi-1D excitonic nature provides a unique 2D platform for investigation of the dynamics of excitons in reduced dimensions and fundamental many-body interactions[11, 18, 19]; this could lead to the development of many novel electronic and optoelectronic devices[1-6, 9-12]. However, on the other hand, the quasi-1D excitonic nature of phosphorene can significantly limit its luminescence quantum yield[11, 12], primarily because of the quenching of mobile excitons caused by the rapid collision of these excitons with local quencher states in the quasi-1D space[20]. Thus, the luminescence quantum yield has been successfully improved through reduction of the defect quenching of excitons in 1D carbon nanotubes[21]. Conversely, instead of acting as exciton quenchers, some local states can function as zero-dimensional (0D) photoluminescent centers, which capture mobile excitons and convert them into photons with a very high radiative relaxation rate. This rate may even lie beyond that of intrinsic 1D excitons[22, 23]. Therefore, the embedding of local 0D-like luminescent states in monolayer phosphorene can lead to brightening of the excitons, which has implications for future optoelectronic and photonics devices.

The deliberate introduction of defects in semiconductors can be realized through irradiation, oxidation, and physical absorption; in particular, the introduction of interfacial surface defect states through substrate engineering has become an important method of manipulating defects in 2D materials, because the large surface-to-volume ratio in atomically thin 2D materials enhances the interfacial interactions with substrates significantly[24, 25]. Specifically, as predicted by previous simulation-based research[26, 27], the highly reactive $sp^3$-like puckered configuration in monolayer phosphorene can lead to the formation of extrinsic point defects, including surface adatoms, which can modify the optical properties of the phosphorene significantly, triggering photon emissions at new wavelengths. In this work, we demonstrate a method of brightening the excitons in monolayer phosphorene via the introduction of luminescent local states to monolayer phosphorene (Figure 1) using PECVD oxide substrate, a commonly used material in photonic devices. Owing to the interfacial luminescent local states, 0D-like localized excitons are formed, resulting in efficient photon emissions at a new wavelength of ~920 nm at elevated temperature in monolayer phosphorene; this contrasts markedly with defect-induced photon emission from TMD semiconductor monolayers that can only be observed at cryogenic temperatures[28]. We find that the luminescence quantum yield of 0D-like localized excitons can be 33.6 times larger than that of the intrinsic quasi-1D free excitons in monolayer phosphorene, mainly caused by the reduction of non-radiative decay rate and possibly the enhanced radiative recombination probability.

**Results and Discussion**

In the experiments presented in this study, mechanical exfoliation technique was used to transfer phosphorene flakes onto PECVD oxide and thermal oxide substrates, respectively, under dry conditions. The phosphorene layer numbers were first estimated based on the contrast observed under an optical microscope (Figure 2a and 2b) and were then quickly and precisely identified using phase-shifting interferometry (PSI) (Figure S1 and S2) and photoluminescence (PL) spectroscopy. Both of these methods were demonstrated as being very robust techniques for determining the layer numbers of exfoliated phosphorene flakes in our previous report[19]. During the PL measurements, the samples were protected at a temperature of −10 °C in a microscope compatible chamber with a slow flow of $N_2$ gas. This was to prevent sample degradation due to photoactivated oxidation caused by aqueous $O_2$[29]. As we have reported previously[19], the low temperature of −10 °C is very important, because the moisture will be frozen in the chamber and this increases the sample lifetime significantly; this allows us to obtain reliable and repeatable measurement data from monolayer phosphorene. Under −10 °C and $N_2$ gas protection, monolayer phosphorene on thermal oxide samples can survive for a few hours, while the monolayer phosphorene samples on PECVD oxide can survive for more than 20 h. The longer lifetime of the monolayer phosphorene samples on PECVD oxide is attributed to the more water-repellent surface of PECVD oxide compared to thermal oxide[30].

The PL spectra measured for the fabricated monolayer phosphorene on thermal oxide substrate sample exhibited one peak at ~720 nm only (labeled the A peak) (Figure 2c); this was attributed to the free excitons of the monolayer phosphorene in accordance with the findings of our previous report[19]. In contrast, the PL spectra measured for the monolayer phosphorene on PECVD oxide substrate exhibited two peaks at ~700 and ~920 nm (labeled the X peak). The peak at ~700 nm was again attributed to the free excitons and, therefore, also labeled the A peak. The slight positional difference of the two A peaks of the examined samples could be due to sample variation and also the slightly different dielectric environments of PECVD oxide and thermal oxide[19]. Interestingly, the X peak at ~920 nm, which occurred in the monolayer phosphorene on PECVD oxide only, had a considerably stronger PL intensity than that of A peak. In order to explore the cause of this new X peak, we ran power-dependent PL measurements. As previously shown by Yu *et al*[31] and Heinz *et al*[32], the integrated PL intensities of localized excitons, trions/excitons, and bi-excitons increase sub-linearly, linearly, and quadratically, respectively, with the growth of excitation powers in a log-log plot[31, 32]. In our measurements, the integrated PL of the low-energy X peak from the monolayer phosphorene sample on PECVD oxide grew sub-linearly with the increased excitation power

($\alpha$ = 0.72, Figure 2d) and the ratio of two PL intensities ($I_X/I_A$) decreased with the increasing of laser power (Figure S3), indicating that the X peak at ~920 nm was from localized excitons.

We found that a strong X PL peak could be reliably observed for all the monolayer phosphorene samples (more than five samples) transferred onto the PECVD oxide, regardless of the type of substrate used under the PECVD oxide. The examined substrates included Au, Si, and transparent quartz. We believe that the localized excitons are related to the interfacial defect states between the monolayer phosphorene and the PECVD oxide. In order to understand the possible mechanisms, we also performed Fourier transform infrared spectroscopy (FTIR) measurements on both the PECVD and thermal oxide substrates. Compared with thermal oxide, the examined PECVD oxide film contained a large number of impurities and surface states (Figure S4), such as SiO bonding groups, which is consistent with previous report[33]. The surface states function as trapping and luminescent centers for the free excitons, which can be also understood from previous simulations based on density functional theory (DFT)[26, 27]. According to reported simulation results[26, 27], extrinsic point defects, including surface adatoms such as O, can modify the electronic properties of monolayer phosphorene significantly. These surface defects can create levels in the bandgap, and such gap states are expected to give rise to recombination lines in luminescence experiments[26, 27]. On the other hand, we also observed a PL emission peak at ~920 nm from partially oxidized monolayer phosphorene samples on thermal oxide substrate. Thus, the X peak at ~920 nm in monolayer phosphorene on PECVD oxide substrate should originate from the oxygen defects in monolayer phosphorene, i.e. horizontal oxygen bridge and/or diagonal oxygen bridge (Figure S5), which is also predicted by DFT calculations[26]. In addition, owing to their large trapping energy[26, 27], the localized exciton emissions in monolayer phosphorene on PECVD oxide can be observed at elevated temperature; in contrast, localized exciton emissions for monolayer TMD semiconductors are only apparent at cryogenic temperatures[28].

Polarization dependent PL is an important technique for the analysis of defect-induced local state symmetry in semiconductors[34, 35]. Based on our angle-dependent PL excitation and emission measurements, we found that the photon emission from the localized excitons (the X peak) in the monolayer phosphorene on PECVD oxide samples has also a linear polarization along the armchair direction of the monolayer phosphorene. This is identical to the case with the free excitons[10]. In Figure 3a, the A states show strong PL excitation polarization dependence, because phosphorene has strongly anisotropic optical absorption[9, 11]. This is consistent with both previous simulation[11] and experimental results[10]. Based on this information, we determined the zigzag direction of the crystalline, in which the PL intensity was at a minimum, and set this direction as the 0° reference of the excitation polarization angle $\theta_1$ and the emission polarization angle $\theta_2$ (Figure 3a, inset). We found that the X states also exhibit strong PL excitation polarization dependence (Figure 3b), which can also be understood as being due to anisotropic absorption in monolayer phosphorene[9, 11, 12]. Then, the polarization of the PL emission for the X states was also measured by fixing $\theta_1$ to 90° (Figure 3c). We found that the PL emissions from the X states in the monolayer phosphorene on PECVD oxide samples have linear polarization along the armchair direction of phosphorene lattice, having the same polarization direction as the quasi-1D free excitons in pristine phosphorene[12]. The fact that both the A and X states (free and localized excitons, respectively) exhibit the same emission polarization along to the armchair direction suggests that the local luminescent states are due to point defects[34, 35]. This further supports the aforementioned conclusion that local luminescent states are highly related to interfacial surface adatom-induced trapping states. On the other hand, by carefully comparing the angle-dependent results in Figure 3a and 3b, although the X exciton show the periodical angle-dependence PL, it does not keep the high anisotropic optical response as A exciton. The ratio of the highest ($\theta_1 = 90°$) and lowest ($\theta_1 = 0°$) PL intensities for X excitons was measured to be ~1.7 (Figure 3b), lower than the ratio value of ~3.5 for A excitons (Figure 3a). We believe the decrease of the anisotropy for the X excitons might come from the defect-induced reduction of local symmetry[26, 36].

Large changes were observed in the PL spectra obtained for different points on the monolayer phosphorene on PECVD oxide substrate, as shown in Figure 4a; this reflects the variable density of the interfacial local states (Figure S6). As the integrated PL intensity of the A peak at ~720 nm decreased slightly, that of the X peak at ~920 nm increased dramatically. Figure 4b shows the variation of the integrated PL intensity for the X state, $\Delta I_X$, as a function of that for the A state, $\Delta I_A$. Here, $\Delta I_X$ is defined as $\Delta I_X = |I_{X1} - I_{X2}|$, where $I_{X1}$ and $I_{X2}$ are the integrated PL intensities for localized exciton states measured at two different locations. $\Delta I_A$ is defined in the same manner but for free excitons. A linear relationship was found between $\Delta I_X$ and $\Delta I_A$, as shown by the dotted points and linearly fitted line in Figure 4b, where the slope of the $\Delta I_X / \Delta I_A$ line is ~12.80. This linear relation provides us with rich information regarding the internal quantum efficiencies of the 0D-like localized excitons and the quasi-1D free excitons in monolayer phosphorene.

Kazunari *et al* proposed an equation that relates $\Delta I_X/\Delta I_A$ to the ratio of the internal quantum yields of free and localized excitons, based on the 1D diffusion-limited exciton contact-quenching mechanism[20, 23], i.e.,

$$\frac{\Delta I_X}{\Delta I_A} \leq \frac{1}{2}\left(\frac{\eta_X}{\eta_A}\right)\left(\frac{E_X}{E_A}\right), \qquad (1)$$

where $\eta_X$ and $\eta_A$ are the luminescence quantum yields of the free and localized excitons, respectively. Thus, a linear relationship between $\Delta I_X$ and $\Delta I_A$ is expected for small $\Delta I_A$, which is consistent with our experimental results (Figure 4b). From the above equation, $\frac{\eta_X}{\eta_A} \geq$ ~33.6 was also derived based on the experimental results, which means that the quantum yield of a single localized exciton in phosphorene is at least ~33.6 times larger than that from a free exciton in monolayer phosphorene.

To understand the origin of the large quantum yield enhancement of the X states, we used time-resolved PL to characterize the carrier lifetimes of the A and X states. The carrier lifetime of the A peak emission in the monolayer phosphorene was measured to be ~211 ± 10 ps (Figure S7), which is consistent with our previously reported value for exfoliated monolayer phosphorene on a thermal oxide substrate[19]. The exact carrier lifetime value of the X peak emission was difficult to be extracted, due to the sensitivity limit of our system. The fact that X peak exhibited strong intensity in PL spectrum but weak intensity in decay curve, indicates a very long carrier lifetime (more than 10 ns) for X states. The localized excitons have significantly longer lifetimes than the free excitons, which is attributed to the highly reduced non-radiative decay rate due to the formation of 0D-like localized excitons[23, 37]. Once a free exciton is trapped by a local X state, its collision possibility with quenching sites will be highly reduced, leading to much longer exciton lifetime and thus higher quantum yield (Figure 4c). On the other hand, the dimensionality modification of the excitons from quasi-1D to 0D can also contribute to the quantum yield enhancement for the localized excitons. The temperature-limited radiative decay rate of 1D excitons ($\propto T^{-1/2}$) also results in lower quantum yield at high temperature, while the temperature-independent radiative decay rate of 0D excitons is free from this limit. This could increase the radiative decay rate and, thus, lead to the enhancement of the quantum yield for 0D-like excitons[23, 37].

**Conclusions**

In conclusion, we have demonstrated a new means of brightening the excitons in monolayer phosphorene to a striking degree, through modifying the dimensionality of excitons from quasi-1D to 0D using PECVD oxide substrate. The interfacial luminescent local states lead to the localization of excitons and trigger efficient photon emissions at a new wavelength of ~920 nm. Using angle-resolved PL measurement, we found that the PL emissions from the localized excitons have linear polarization along the armchair direction of phosphorene lattice, having the same polarization direction as the quasi-1D free excitons in pristine monolayer phosphorene, suggesting that the local luminescent states are due to point defects. Owing the large trapping energy, localized excitons in monolayer phosphorene were observed at elevated temperature. This contrasts markedly with the defect-induced photon emission from TMD semiconductor monolayers is only apparent at cryogenic temperatures. More importantly, the luminescence quantum yield of the 0D-like localized excitons was measured to be at least 33.6 times larger than that of the intrinsic quasi-1D free excitons in monolayer phosphorene. Our findings open up new avenues for the development of novel photonic devices based on monolayer phosphorene. Moreover, 2D phosphorene with quasi-1D free excitons and 0D-like localized excitons provides a unique platform to explore the fundamental interactions in the 2D-1D-0D hybrid systems.

**Methods**

Mono- and few-layer phosphorene samples were produced by mechanical exfoliation from black phosphorus crystal (Smart Element) and transferred onto PECVD oxide (223 nm)/Au substrate and Si substrate capped with 275 nm of thermal oxide, respectively. PECVD oxide was deposited at 200 ºC with $SiH_4$ and $N_2O$ as precursors, using the Plasmalab 100 dual frequency system. Monolayer phosphorene samples were put into a Linkam THMS 600 chamber and the temperature was set as during the PL measurements. All PL spectra, power-dependent PL and polarization-dependent PL measurements were conducted at a temperature of $-10\ ℃$, in a microscope compatible chamber (Linkam THMS 600), using a 532nm Nd:YAG laser as the excitation source. Our PL system (T64000) has two liquid nitrogen cooled detectors, one charge-coupled device (CCD) detector and one InGaAs detector. The time-resolved decay curves were measured using a time-correlated single photon counting (TCSPC) system, using a 300 fs pulse laser (20.8 MHz repetition rate) at a wavelength of 522 nm.

**Author Contributions**

Y. R. L. initialized the project; R. J. X, Y. W. M, and J. J. P. prepared the samples; J. Y. fabricated the PECVD oxide substrates; Y. W. M and J. Y. conducted the PL measurements; R. J. X.

analyze the data and prepared the manuscript draft; F. W. built the optical characterization setup; All authors have contributed to the writing of manuscript.

**Acknowledgements**

We want to thank Professor C. Jagadish, and Professor B. L.-Davies for their facility support. We would like to acknowledge financial support from the ANU PhD scholarship, the China Research Council PhD scholarship, the Australian Research Council (grant number DE140100805), and the ANU Major Equipment Committee.

**Competing Financial Interests**

The authors declare that they have no competing financial interests.

**FIGURE CAPTIONS**

**Figure 1 | Schematic of exciton dynamics in monolayer phosphorene with a 0D-like local luminescent state. a,** Energy band diagram of free and localized excitons in monolayer phosphorene. The 0D-like local luminescent state captures a free exciton and confine it to a lower energy state. **b,** A free exciton in monolayer phosphorene diffusively moves along the armchair axis, converts to a 0D-like exciton at the local state and emits brighter PL than quasi-1D excitons.

**Figure 2 | Photoluminescence (PL) spectra of monolayer phosphorene on PECVD oxide/Au and thermal oxide/Si substrate. a-b,** Optical microscope images of mechanically exfoliated monolayer phosphorene flakes on PECVD oxide (a) and thermal oxide (b) substrates. Inset shows the schematic structure of corresponding device. **c,** Measured PL spectra of the monolayer phosphorene on PECVD oxide/Au and thermal oxide/Si substrates. The measured PL spectra from substrate background are also plotted as comparison. The peaks from free excitons are labelled as A peak, where is at ~720 nm for thermal oxide/Si substrate and at ~700 nm for PECVD oxide/Au substrate. The second low-energy PL peak at ~920 nm from the monolayer phosphorene on PECVD oxide is labelled as X peak. **d,** Integrated PL of the low-energy X peak from the monolayer phosphorene sample on PECVD oxde/Au substrate. The sub-linearly growth of integrated PL with the increased excitation power ($\alpha = 0.72$), indicating that the X peak at ~920 nm is from localized excitons.

**Figure 3 | Angle-resolved PL measurements for monolayer phosphorene on PECVD oxide/Au substrate. a,** Measured excitation polarization dependence of PL peak intensities from free excitons in monolayer phosphorene on PECVD oxide/Au substrate. Inset: schematic plot showing top view of phosphorene lattice structure and coordinates for incident excitation polarization angle $\theta_1$ and PL emission polarization angle $\theta_2$. The incident excitation polarization angle ($\theta_1$) is controlled by an angle-variable half-wave plate, and the PL emission polarization angle ($\theta_2$) is characterized by inserting an angle-variable polarizer in front of the detector. **b,** Measured excitation polarization dependence of PL peak intensities from localized excitons in monolayer phosphorene on PECVD oxide/Au substrate. For the measurements in (a) and (b), the polarizer in front of the detector was removed. **c,** Measured emission polarization dependence of PL intensities for localized excitons in monolayer phosphorene on PECVD oxide/Au substrate, with fixed excitation angle of 90°.

**Figure 4 | Relationship of luminescence intensities from free and localized excitons in monolayer phosphorene on PECVD oxide/Au substrate. a,** Measured PL spectra from different locations in the monolayer phosphorene on PECVD oxide/Au substrate shown in Figure 2a. **b,** Integrated PL intensity of variation for localized excitons, $\Delta I_X$, as a function of that for free exciton state, $\Delta I_A$. The integrated PL intensities are evaluated by peak decomposition procedures, where the free exciton and localized exciton peaks are fitted by a Gaussian function. **c,** Schematic of exciton diffusion and successive trapping by local quenching sites or local luminescent sites.

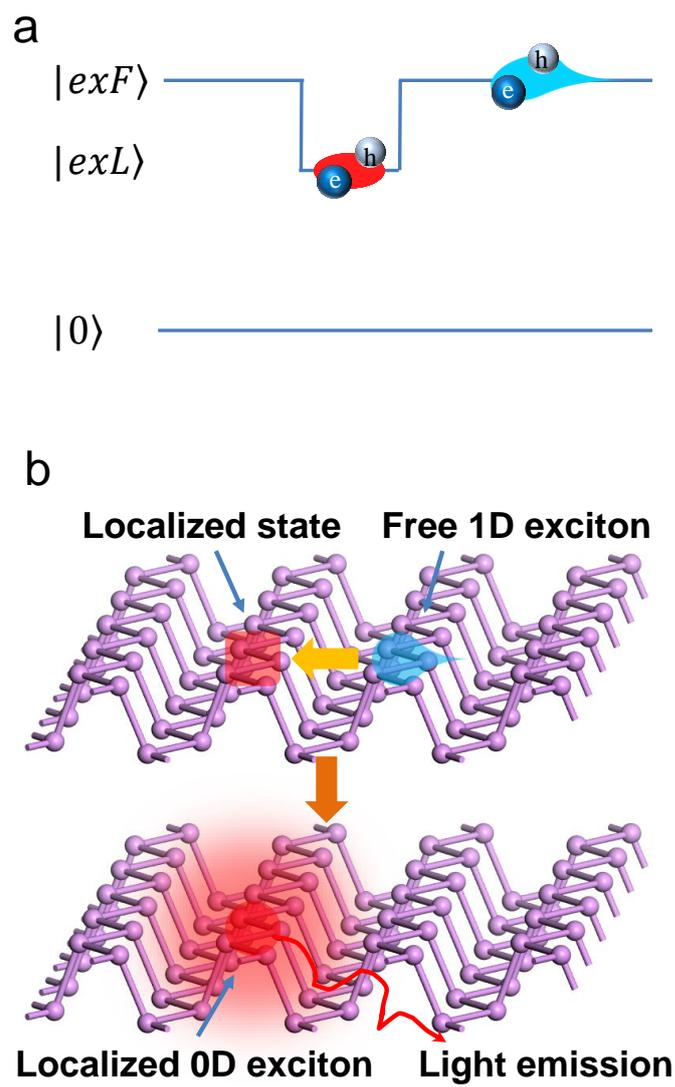

Figure 1

a 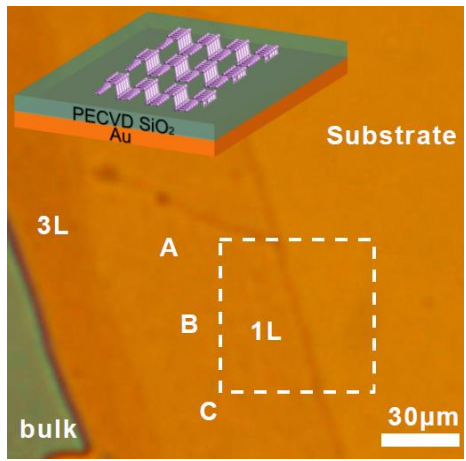 b 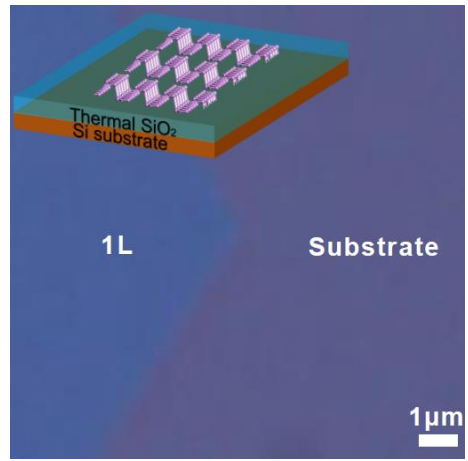

c 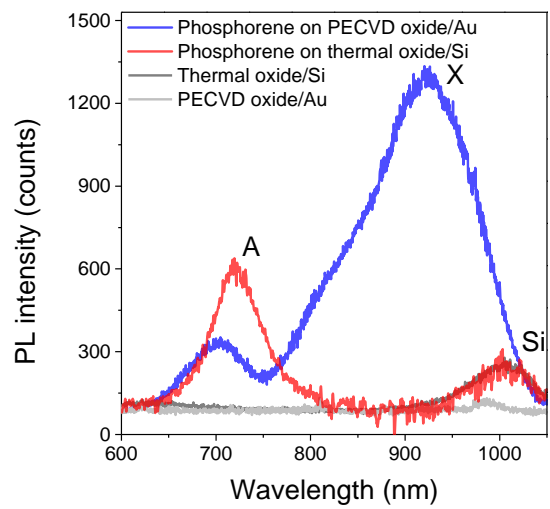 d 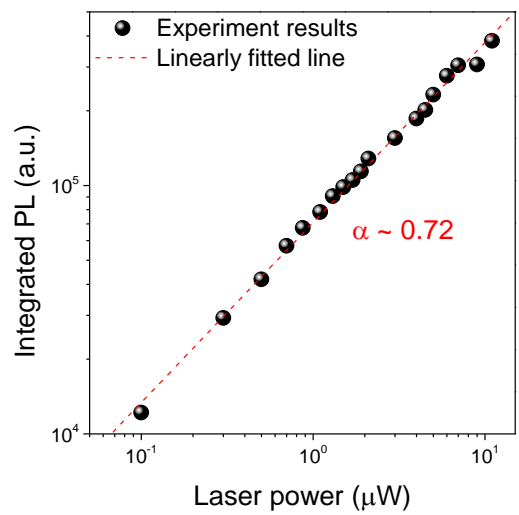

Figure 2

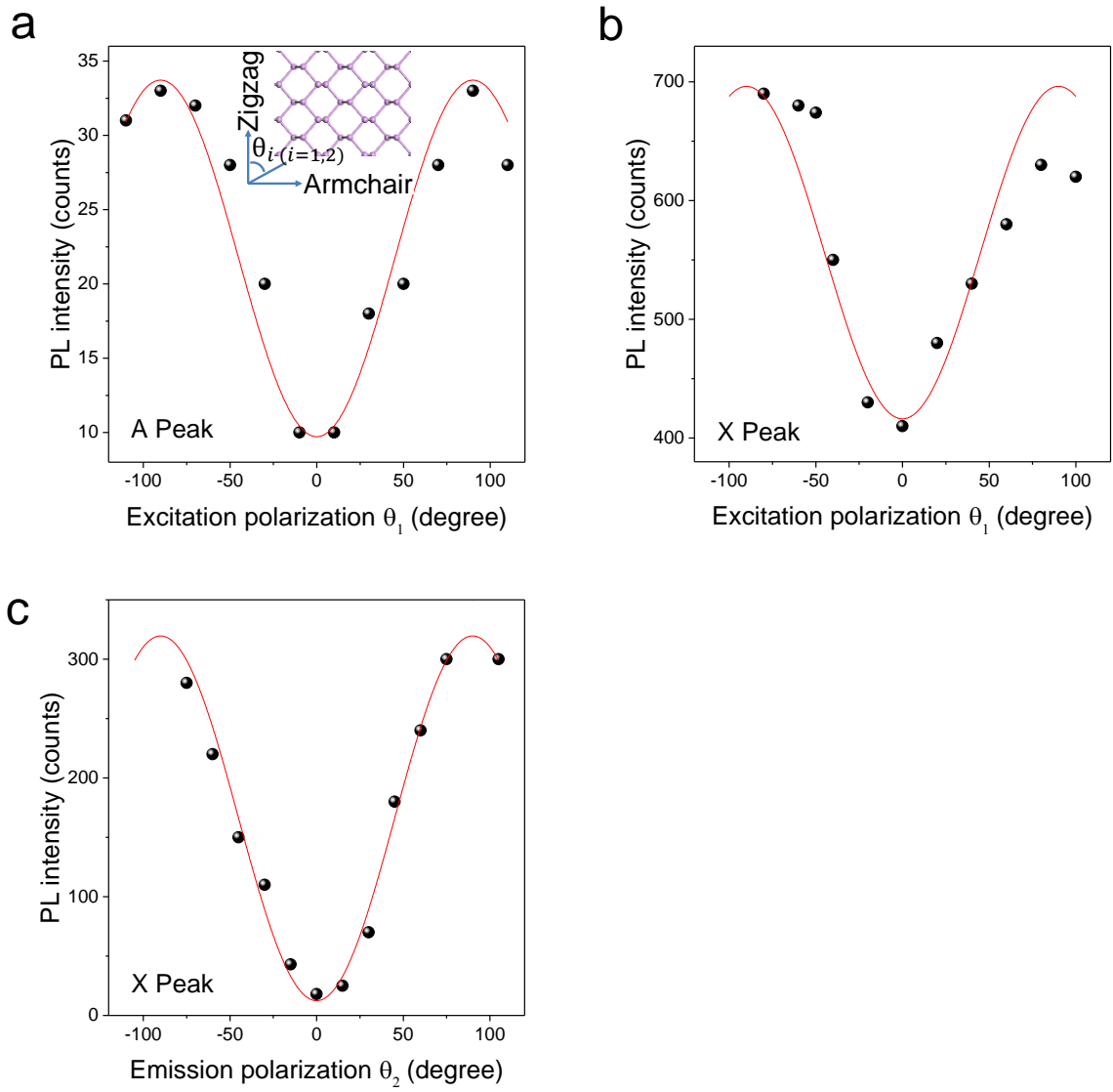

Figure 3

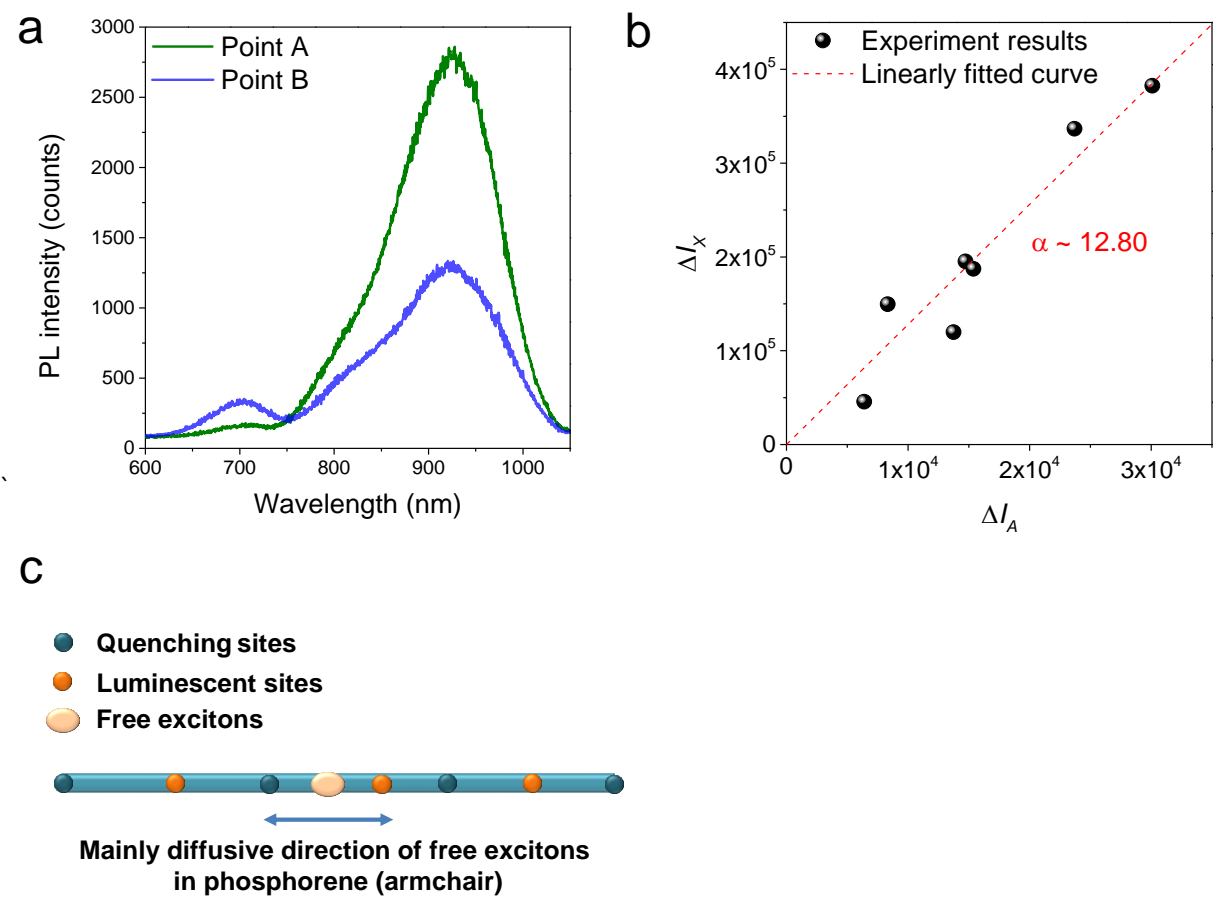

Figure 4

# Supplementary Information for

# Exciton Brightening in Monolayer Phosphorene via Dimensionality Modification

Renjing Xu,[1,†] Jiong Yang,[1,†] Ye Win Myint,[1] Jiajie Pei,[1,3] Han Yan,[1] Fan Wang,[2] and Yuerui Lu[1*]

[1]Research School of Engineering, College of Engineering and Computer Science, the Australian National University, Canberra, ACT, 2601, Australia

[2]ARC Centre for Nanoscale BioPhotonics (CNBP), Department of Physics and Astronomy Faculty of Science, Macquarie University, Sydney, NSW, 2109, Australia

[3]School of Mechanical Engineering, Beijing Institute of Technology, Beijing, 100081, China

[†] These authors contributed equally to this work

**\*** To whom correspondence should be addressed: Yuerui Lu (yuerui.lu@anu.edu.au)

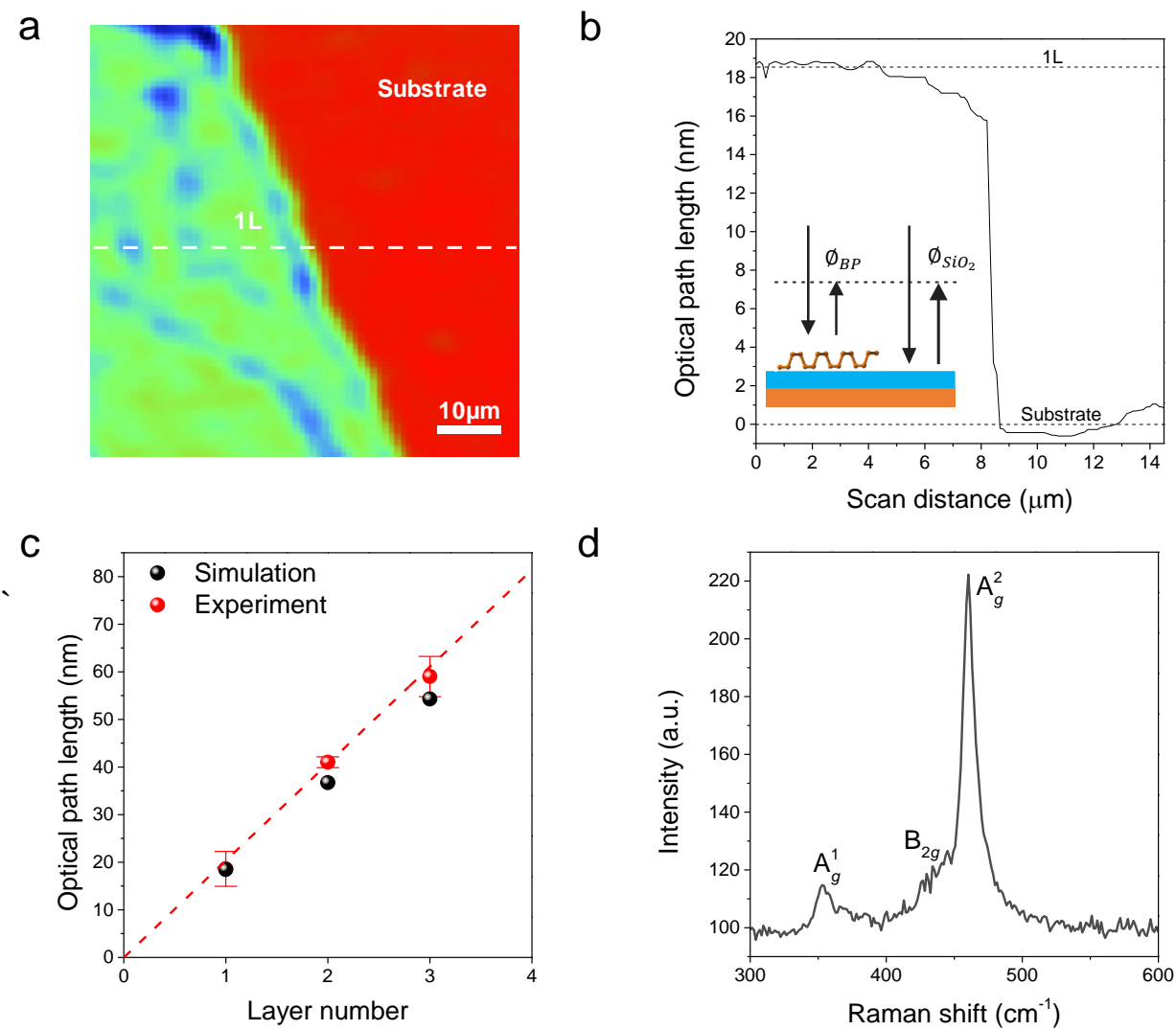

**Figure S1 | Characterization of monolayer phosphorene on PECVD oxide/Au substrate.**

**a,** Phase-shifting interferometry (PSI) image of the monolayer phosphorene from the box enclosed by the dash line in Figure 2a. **b,** PSI measured optical path length (OPL) values versus scan position for monolayer phosphorene along the dash line in (a). The measured optical path length (OPL) value of the sample on thermal oxide/Si substrate is 18.6 nm, indicating monolayer, according to statistical data in (c). Inset: schematic plot indicating the PSI measured phase shifts of the reflected light from the phosphorene flake ($\phi_{BP}$) and the SiO$_2$ substrate ($\phi_{SiO_2}$). **c,** Statistical data of the OPL experimental values from PSI for 1–3L phosphorene samples on PECVD oxide (223 nm)/Au substrate. For each layer number of phosphorene, at least two different samples were characterized for the statistical measurements. Theoretical simulation data is also plotted, for comparison. The red dash line is the linear trend for statistical data measured with the PSI system. **d,** Raman spectrum of monolayer phosphorene on PECVD oxide/Au substrate.

*Numerical Simulation*. Stanford Stratified Structure Solver (S4)[1] was used to calculate the phase delay. The method numerically solves Maxwell's equations in multiple layers of structured materials by expanding the field in the Fourier-space.

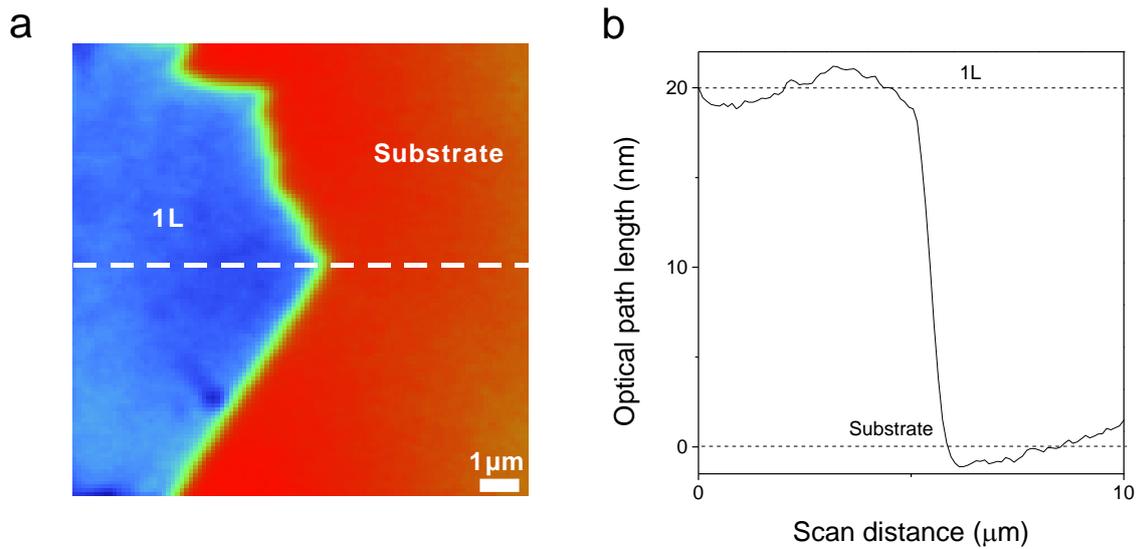

**Figure S2 | Characterization of monolayer phosphorene on thermal oxide/Si substrate. a,** PSI image of the mono-phosphorene in Figure 2b. **b,** PSI measured OPL values versus scan position for monolayer phosphorene along the dash line in (a). The measured optical path length (OPL) value of the sample on thermal oxide/Si substrate is 19.9 nm, indicating mono-layer, according to our previous report[2].

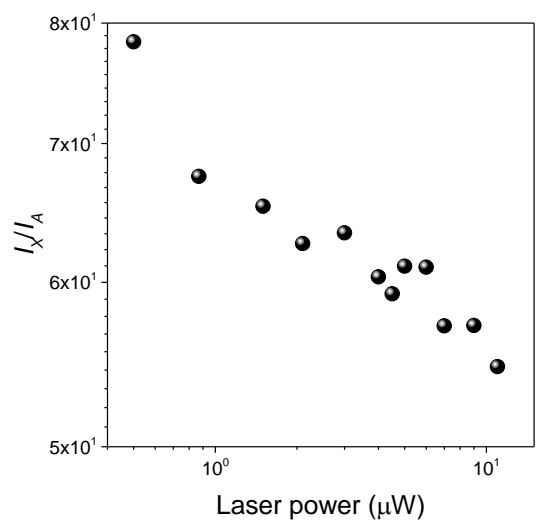

**Figure S3** | Measured ratio of two PL intensities ($I_X/I_A$) as a function of laser power.

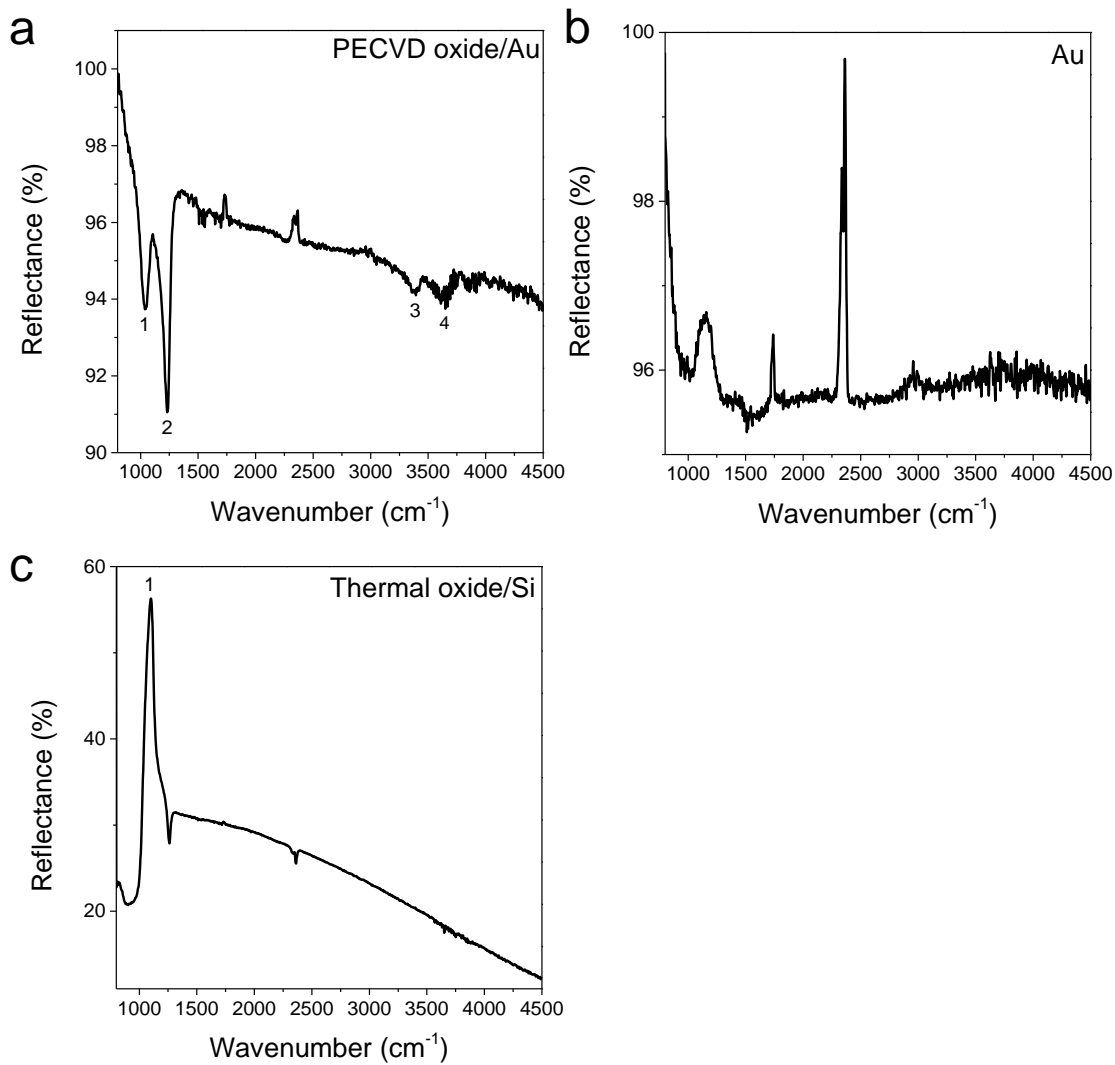

**Figure S4 | Fourier transform infared spectroscope (FTIR) measurements on PECVD oxide and thermal oxide. a,** FTIR spectrum for PECVD oxide/Au substrate. Frequencies of various impurities atom modes due to SiO (peak 1 at 1020 cm$^{-1}$ and peak 2 at 1220 cm$^{-1}$), NH (peak 3 at 3350 cm$^{-1}$) and OH (peak 4 at 3640 cm$^{-1}$) bonding groups in PECVD oxide were detected[3, 4]. **b,** FTIR spectrum for Au. **c,** FTIR spectrum for thermal oxide/Si substrate. The Si-O symmetric stretch mode is at wavenumber of 1090 cm$^{-1}$ (peak 1) [3, 4].

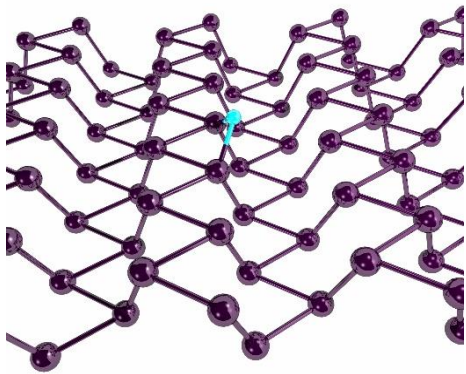 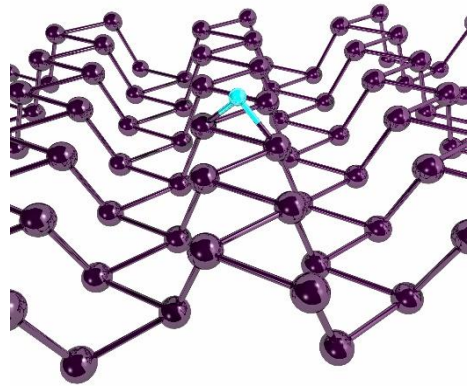

**Figure S5 |** Schematic of Horizontal bridge (a) and diagonal bridge (b) configurations for oxygen defects[5].

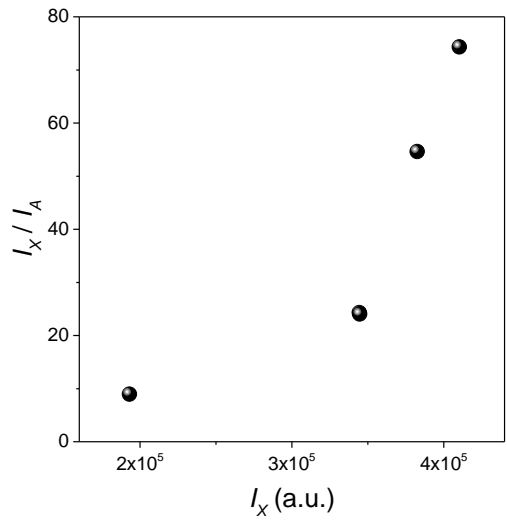

**Figure S6** | Measured ratio of two PL intensities ($I_X/I_A$) as a function of $I_X$.

Here, if we simply assume that the increase of $I_X$ is owning to the increase of defect concentration, then the growth of the defect concentration will directly result in higher $I_X/I_A$ ratio as the $I_A$ would decrease and $I_X$ would increase in the meantime. We could clearly see this trend from our measured data (Figure S6).

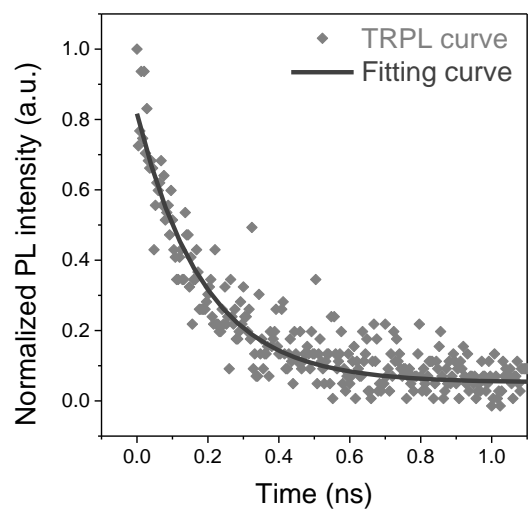

**Figure S7 | Lifetime decay curve of the free exciton peak in monolayer phosphorene on PECVD/Au substrate.** PL decay profile of the exciton peak in monolayer phosphorene was measured, by using a pulse 522 nm (frequency doubled) excitation laser at a laser power of 1.15 μW and a time-correlated single photon counting system (TCSPC). A lifetime of ~211 ± 10 ps was extracted from the PL decay curve by exponential fitting. The grey solid line curve is the exponential fitting of the experimental data.